\documentclass[9pt,twocolumn,twoside]{pnas-new}

\templatetype{pnasresearcharticle} 

\title{Hydrodynamics of Random-Organizing Hyperuniform Fluids}

\author[a]{Qun-Li Lei}
\author[a,1]{Ran Ni}

\affil[a]{School of Chemical and Biomedical Engineering, Nanyang Technological University, 62 Nanyang Drive, 637459, Singapore}

\leadauthor{Qun-Li Lei} 

\significancestatement{Recently, an exotic hyperuniform fluid state was found in a number of non-equilibrium systems, which shows the promise for fabrication of novel {lifelike} functional materials capable of self-healing and self-adapting. However, the general mechanism of fluidic hyperuniformity (HU) and its fundamental difference from equilibrium fluids remain unclear, which hinders the rational design of non-equilibrium hyperuniform fluids. In this work, we propose a non-equilibrium hard-sphere model of hyperuniform fluids and formulate a full hydrodynamic description based on the generalized Navier-Stokes equations. Our theory reveals the mechanism of fluidic HU in a simple and intuitive way, which is {confirmed} by simulating a realistic active spinner system. }

\authorcontributions{Author contributions: Q.-L. L. and R. N. designed the research; Q.-L. L. performed the research and analyzed the data; Q.-L.L. and R.N. wrote the paper.}
\authordeclaration{The authors declare no conflict of interest.}
\correspondingauthor{\textsuperscript{1}To whom correspondence should be addressed. E-mail: r.ni@ntu.edu.sg}


\begin{abstract}
Disordered hyperuniform structures are locally random while uniform like crystals at large length scales. Recently, an exotic hyperuniform fluid state was found in several non-equilibrium systems, while the underlying physics remains unknown.
In this work, we propose a non-equilibrium (driven-dissipative) hard-sphere model and formulate a hydrodynamic theory based on Navier-Stokes equations to uncover the general mechanism of the fluidic hyperuniformity (HU).  At a fixed density, this model system undergoes a smooth transition from an absorbing state to an active hyperuniform fluid, then to the equilibrium fluid by changing the dissipation strength. We study the criticality of the absorbing phase transition. We find that the origin of fluidic HU can be understood as the damping of a stochastic harmonic oscillator in $q$ space, which indicates that the suppressed long-wavelength density fluctuation in the hyperuniform fluid can exhibit as either acoustic (resonance) mode or diffusive (overdamped) mode. Importantly, our theory reveals that the damping dissipation and active reciprocal interaction (driving) are two ingredients for fluidic HU. Based on this principle, we further demonstrate how to realize the fluidic HU in an experimentally accessible active spinner system and discuss the possible realization in other systems.
\end{abstract}

\dates{This manuscript was compiled on \today}
\doi{\url{www.pnas.org/cgi/doi/10.1073/pnas.XXXXXXXXXX}}

\begin{document}

\maketitle
\thispagestyle{firststyle}
\ifthenelse{\boolean{shortarticle}}{\ifthenelse{\boolean{singlecolumn}}{\abscontentformatted}{\abscontent}}{}


\dropcap{H}yperuniform structures are characterized by  vanishing long-wavelength density fluctuations with the structure factor $S(q \rightarrow 0) = 0$~\cite{torquato2003local,torquato2018PR}, which include perfectly ordered structures, such as athermal crystals or quasi-crystals. In past decades, an increasing number of disordered structures were also found to exhibit hyperuniformity (HU), e.g., maximally random jammed states~\cite{donevprl2005},  animal photoreceptor cell patterns~\cite{kram2010avian,yellott1983}, { low-discrepancy sequences~\cite{kuipers2012uniform},  eigenvalues of random matrices~\cite{ginibre1965}, early universe fluctuations~\cite{gabrielli2002glass}} etc.~\cite{torquato2018PR}. These disorder structures have shown even better properties than crystals usually exclusively have, like isotropic photonic bandgaps opened at low dielectric contrast~\cite{florescu2009designer,man2013isotropic,man2013photonic}, abnormal transparency~\cite{batten2008classical,leseur2016high}  and also has many applications in graphic anti-aliasing~\cite{cook1986} and quasi-Monte Carlo sampling~\cite{a2007stochastic}.
Disordered HU implies a long-range direct correlation function in particle systems~\cite{torquato2003local,torquato2018PR}. 
In equilibrium, this requires delicately designed long-range interactions~\cite{batten2008classical,lomba2017disordered}, which are challenging for experimental realization.
Recently, HU has been found in some dynamic states of non-equilibrium systems, including the non-ergodic dynamic critical states~\cite{hexner2015,tjhung2015,hexner2017enhanced,weijs2015emergent,schrenk2015,wang2018h} and  ergodic fluid states~\cite{hexner2017noise,lei2019}. The latter shows the promise of realizing novel flowing functional materials, e.g., photonic fluids~\cite{bi2016tissue}.
Current studies of dynamic HU mainly focus on time-discrete random-organizing models, in which particles interacting with their neighbours perform activated random displacement~\cite{corte2008random,hexner2015,tjhung2015,hexner2017enhanced,hexner2017noise,ma2019hyperuniformity}.
However, the general mechanism of the non-equilibrium hyperuniform fluids and the fundamental difference from equilibrium fluids remains unknown, let alone a full hydrodynamic description of these exotic fluids. To address these questions, in this work, we propose a non-equilibrium hard-sphere fluid model system, which can smoothly transform from an equilibrium simple fluid to a non-equilibrium hyperuniform fluid. We construct a hydrodynamic theory for the hyperuniform fluids based on the generalized Navier-Stokes equations. The theory reveals the mechanism of fluidic HU intuitively through a damped stochastic harmonic oscillator in $q$ space. Based on the analysis, we obtain two important ingredients for the fluidic HU and further show the realization in an experimentally accessible 2D active spinner system. Our findings suggest possible ways to realize fluidic photonic materials that can heal from damage, and their optical properties can be externally controlled~\cite{bachelard2017emergence}.

\section*{Results}
\subsection*{Model}
\begin{figure}[!htb] 
\centering
\begin{tabular}{c}
	\resizebox{80mm}{!}{\includegraphics[trim=0.0in 0.0in 0.0in 0.0in]{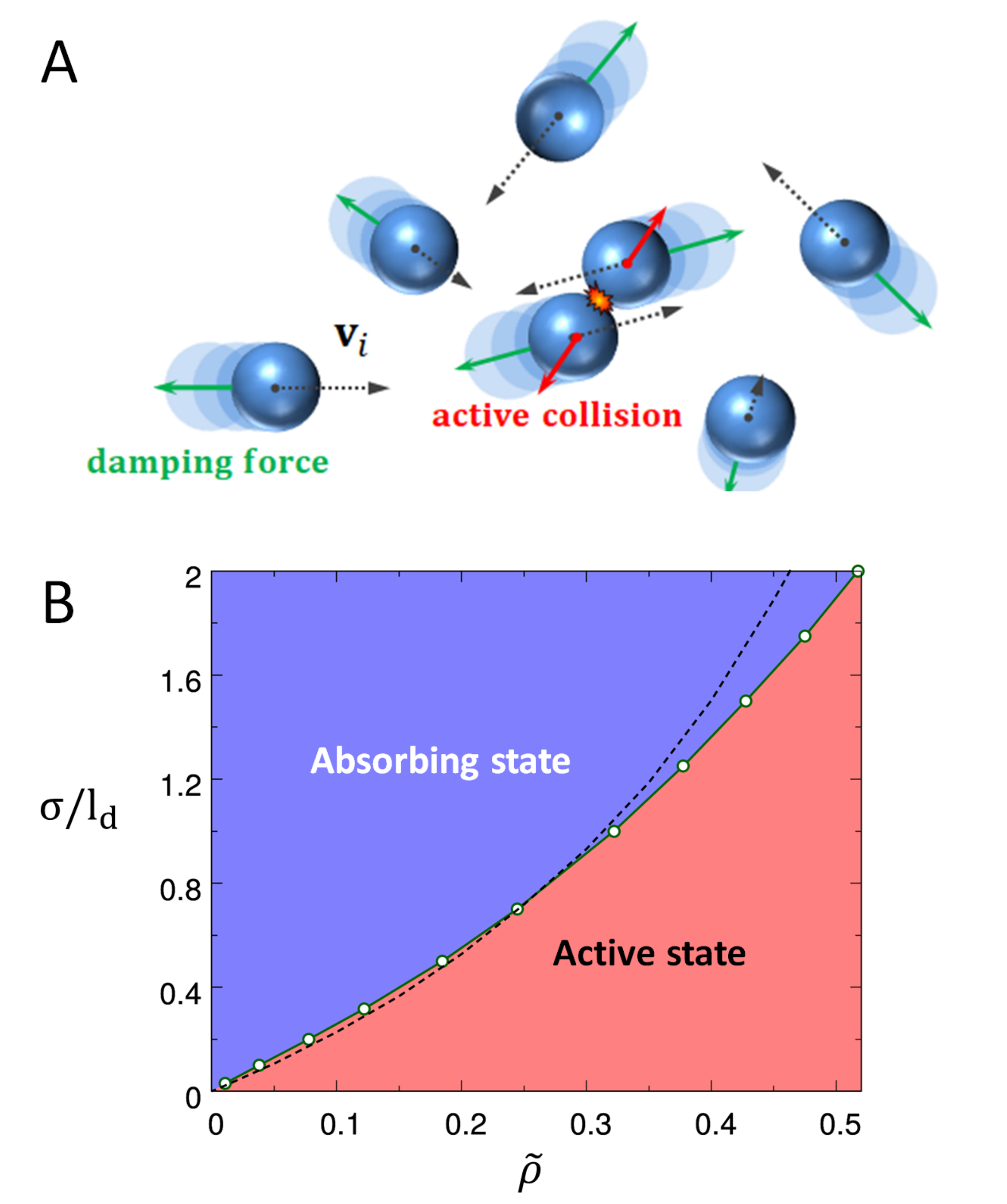} }
\end{tabular}
\caption{(\emph{A}) Schematic of the random-organizing hard sphere fluid. (\emph{B}) Phase diagram of 2D system in the representation of the reduced density $\tilde{\rho}$ and the inverse dissipation length $\sigma/l_d$. The dashed line shows mean free path $l_m^{-1}(\tilde{\rho})$.}
\label{Fig1}
\end{figure}
We first consider a minimal model consisting of $N$ hard spheres (See Fig. \ref{Fig1}A). Each particle has a mass $m$, a diameter $\sigma$ and a random initial velocity. Particles undergo active collision with reciprocal center-to-center impulsions, in which an additional kinetic energy $\Delta E$ is injected through each collision between two particles. The kinetic energy is then dissipated through the (frictional) damping, and the equation motion of particle $i$ between two consecutive collisions is
\begin{eqnarray} 
m\frac{d {\mathbf v_i}(t)}{d t}=-\gamma {\mathbf v_i}(t),
\end{eqnarray}
with $\gamma$ the damping coefficient. This model is a generalization of previous time-discrete random-organizing models~\cite{hexner2015,tjhung2015,hexner2017enhanced,
hexner2017noise,ma2019hyperuniformity} by incorporating the inertia of the particles and it has two characteristic lengths: i) the mean free path $l_m$, which at low density is proportional to ${\sigma }/{\tilde{\rho}}$, with $\tilde{\rho}=N\sigma^d/V$ the dimensionless particle number density, $V$ the volume of the system and $d$ the dimension of the system; ii) the dissipation length $l_d$, i.e., the typical distance that an isolated activated particle can travel, 
\begin{eqnarray} 
l_{d} \equiv {\sqrt{ m \Delta E} }/{{\gamma} },
\end{eqnarray}
which represents the strength of inertia of the particles. Here we assume two particles collide with vanishing approaching velocities and $\sqrt{\Delta E/m}$ is the typical activated velocity to each particle. As our system only has two energy~(time)~-related quantities, i.e., $\Delta E$ and $\gamma$, the typical dissipation time associated with $\gamma$ can be written as $\tau_d = m/\gamma=(l_d/\sigma) \tau_0$ by assuming $\tau_0 \equiv \sqrt{m\sigma^2/\Delta E}$ the unit of time. Therefore, $l_m$ and $l_d$ uniquely define the state of the system. We devise a new event-driven algorithm to simulate this underdamped system in both 2D and 3D with full periodic boundary conditions (see Method for the simulation details). As the conclusion we obtain is independent of dimensionality, in the following, we focus on the 2D systems and leave the results of 3D systems in \emph{SI Appendix}.

\subsection*{Absorbing Phase Transition}
In our model with random initial particle velocities, when $l_{d} \ll l_m$, a single active collision on average induces less than one further collision, which causes the exponential decay of activation events as a function of time. The system thus falls into an absorbing state with vanishing particle velocity, in some sense analogous to the liquid-solid transition (see \emph{SI Appendix} {Movie} S1). On the contrary, when $l_{d} \gg l_m $, a single activation can trigger chain reactions, leaving the system in a dynamically active state {(see \emph{SI Appendix} {Movie} S2, S3)}. Similar absorbing phase transitions have been reported in many other systems~\cite{corte2008random,regev2015,royer2015precisely,
mangan2008reversible}, which have deep connections with the celebrated self-organized criticality~\cite{jensen1998self}. { In this work, we focus on the low density regime, to avoid the possible crystallization}. In Fig.~\ref{Fig1}B, we plot the phase diagram of the 2D system in the representation of $\sigma/l_d$ and $\tilde{\rho}$. The solid line with open symbols shows the phase boundary of system obtained from simulations, and the dashed line shows  $\sigma/l_m$ as a function of $\tilde{\rho}$. Two lines are rather close to each other especially at small $\tilde{\rho}$, indicating the transition occurs round $l_{d} \simeq l_m $.  In \emph{SI Appendix}, Fig.~S1-2, by using the finite-size scaling analysis~\cite{henkel2008non,rossi2000}, we investigate the criticality of this transition in 2D systems with weak inertia ($l_{d} =\sqrt{10}\sigma$) and strong inertia ($l_{d} =10^4 \sigma$). We summarize the obtained critical exponents in Table \ref{Tab_1} and compare them with those from the universality of directed percolation (DP)~\cite{henkel2008non} and conserved directed percolation (CDP)~\cite{manna1991}. We find the transition belongs to CDP for systems with small inertia and relative high critical density $\tilde{\rho}_c=0.12192$. Increasing the inertia to $l_d=10^4\sigma$ makes the system approach the dilute limit ($\tilde{\rho}_c=3.2350\times 10^{-5}$) and induces a crossover from CDP to its long-range mean-field scenario due to the corresponding increase of $l_m$. Here, $l_m$ acts as the effective ``interaction range" in the system, which diverges as the system approaches the dilute limit. Similar results for 3D systems can be found in \emph{SI Appendix}, Fig.~S3-4 and Table S1.
\begin{table}[h!]
\begin{tabular}{c c c c c}
\hline \hline \\[-2.0ex]
 ~~& ~~~~$\beta$~~~~ & ~~~~$\alpha$~~~~ & ~~~~$\nu^*_{\perp}$~~~~ & ~~~~$z^*$~~~~  \\[0.5ex]
\hline 
\\[-1.5ex]
 ~DP~~ & 0.583(3) & 0.450(1) & 0.733(8) & 1.766(2)  \\ [0.5ex]
 
 CDP~~ & 0.64(1)  & 0.52(1) & 0.80(2)  & 1.53(2)   \\ [0.5ex]
 
Mean-field (long-range)~~~ & 1 & 1 & 1 & 1 \\ [0.5ex]

$\tilde{\rho}_c=0.12192(2)$               & 0.65(2) & 0.54(2) & 0.84(3) & 1.49(3)  \\ [0.5ex]
$\tilde{\rho}_c=3.2350(5)\times 10^{-5}$  & 1.02(3) & 1.03(3) & 0.98(3) & 1.00(2)  \\ [0.5ex]
\hline \hline
\end{tabular}
\caption{ Comparison of critical exponents measured in  2D systems with those from universality classes of DP, CDP \cite{henkel2008non,lee2013comment} and their long-range mean-field values. } \label{Tab_1}
\end{table}

\begin{figure*}[!htb] 
\centering
\begin{tabular}{c}
	\resizebox{180mm}{!}{\includegraphics[trim=0.0in 0.0in 0.0in 0.0in]{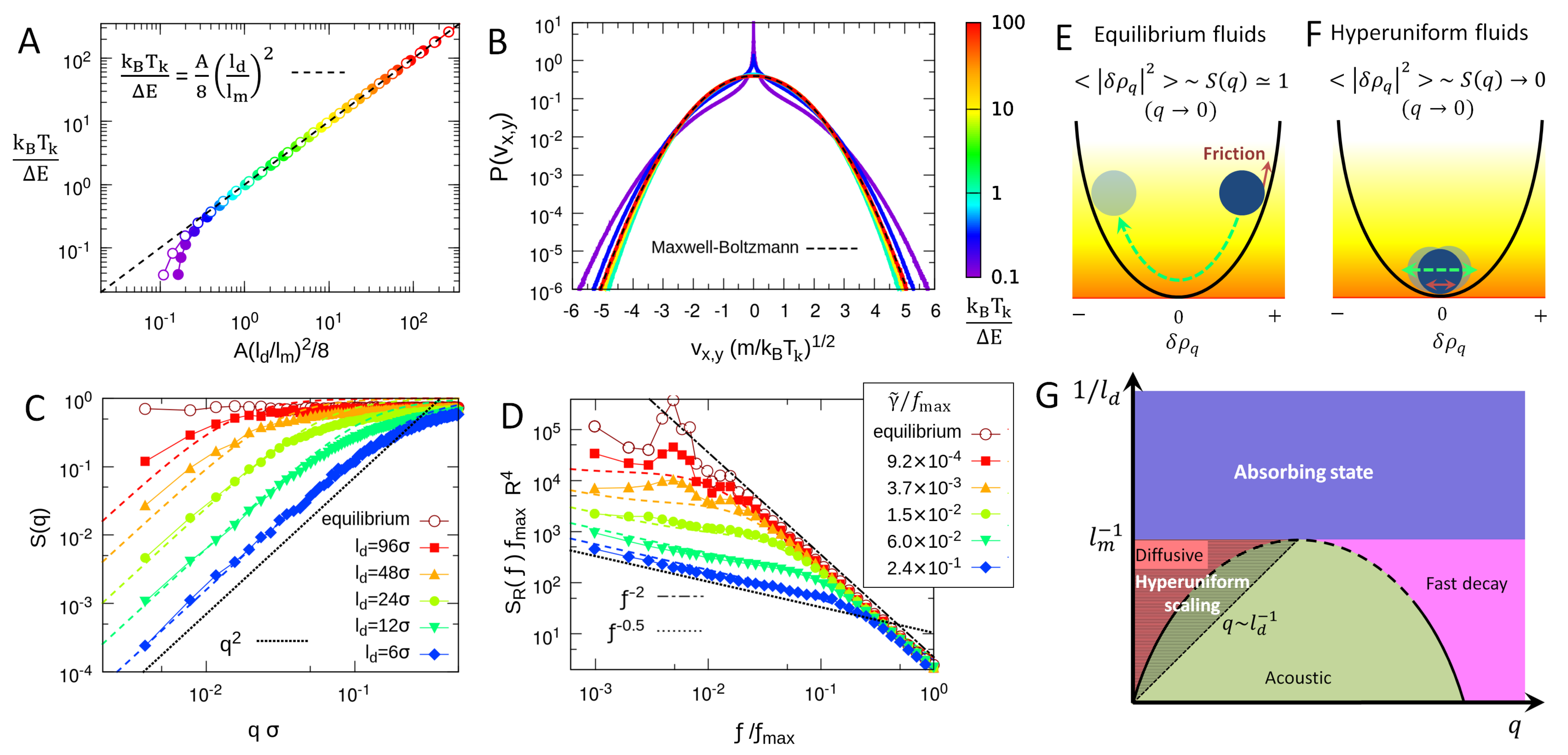} }
\end{tabular}
\caption{(\emph{A}) Kinetic temperature $T_{k}$  as a function of $\frac{A}{ 8}\left(\frac{l_d}{l_m}\right)^2$ for systems with fixed $l_m=4.4\sigma$ ($\tilde{\rho}=0.1$, solid symbols) and fixed $l_d=160\sigma$ (open symbols).~ (\emph{B}) Velocity distributions of particles along $x$ and $y$ directions at different ${k_BT_{k}}/{\Delta E}$ with fixed $l_m=4.4\sigma$. The dashed line indicates the Maxwell-Boltzmann distribution at temperature $T_{k}$. (\emph{C}) Structure factor $S({q})$ for systems with different $l_d$ at $l_m=4.4\sigma$. The dotted line indicates the $q^{2}$ hyperuniform scaling. The dashed lines show the theoretical predictions of Eq.~(\ref{Sk_final}) with no adjustable coefficient. (\emph{D}) Power spectrum of local density fluctuation $S_R(f)$ for the same systems in {\emph{C}} distinguished by  $\tilde{\gamma}/f_{max}$. The dashed lines show the theoretical predictions multiplied by an adjustable coefficient $0.62$. \emph{A, B, C, D} share the same color coding of $T_{k}/\Delta E$ except for the equilibrium fluid ($l_d=\infty$ and $\tilde{\gamma}/f_{max}=0$) shown as open symbols. In \emph{C, D}, $N=512^2$ for all cases. (\emph{E, F}) The density fluctuation $\delta \rho_q$ in the representation of damped stochastic harmonic oscillators for ({\emph E}) equilibrium fluids  and ({\emph F}) non-equilibrium hyperuniform fluids. {  Here the dashed and solid arrows only roughly indicate the particle motion and the friction force, respectively.} ({\emph G})  Schematic diagram of the density modes of the system.}
\label{Fig2}
\end{figure*} 

\subsection*{From Equilibrium Simple Fluids to Non-Equilibrium Hyperuniform Fluids}
Our model is a driven-dissipative system. In the active state, the power of energy injection and dissipation per particle at the mean-field level can be written as
\begin{eqnarray} 
W_{driv}\simeq  f_a \Delta E,~~~~~~~~~~~~~
W_{disp}\simeq \overline{v}^2{\gamma}, \label{W_disp}
\end{eqnarray}
respectively, where $f_a \equiv {\overline{v}}/2{l_m} $ is the average collision frequency per particle and $\overline{v}$ is the average particle speed. Since the dissipation is a quadratic function of $\overline{v}$, a driving-dissipation balanced state with $W_{disp} = W_{driv}$ can be reached, which leads to the kinetic temperature of the system:
\begin{eqnarray} 
 k_B T_{k} \equiv \frac{m\overline{{v}^2} }{d}\simeq \frac{m\overline{v}^2 }{d} \simeq  \frac{ \Delta E  }{4d}\left(\frac{l_d}{l_m}\right)^2.
\end{eqnarray}
In Fig.~\ref{Fig2}A, we plot $k_B T_{k}/\Delta E$ measured in the simulation for 2D systems with different $l_d$ at $l_m=4.4\sigma$ ($\tilde{\rho}=0.1$) (solid symbol) and  the systems with different $l_m$ at fixed $l_d=160\sigma$ (open symbols). We find the two sets of data collapsing onto a single curve when $l_d\gg l_m$, which can be written as $\frac{k_B T_{k}}{\Delta E}=\frac{A  }{ 8}\left(\frac{l_d}{l_m}\right)^2$, with the pre-factor $A=1.56$. The distributions of particle velocity in $x, y$ directions, i.e., $P(v_x)$, $P(v_y)$, for systems at $l_m=4.4\sigma$ with different $k_B T_{k}/\Delta E$ are shown in Fig.~\ref{Fig2}B. We find that $P(v_x)$ and $P(v_y)$ are statistically indistinguishable from each other. 
With increasing $k_B T_{k}/\Delta E$, the velocity distribution gradually changes from non-Gaussian distribution (a signature of non-equilibrium fluids) to the Maxwell-Boltzmann (Gaussian) distribution: that is, $P(v_i)= \left( \frac{m}{2\pi k_BT_{k}} \right)^{1/2} \exp\left[ \frac{-mv_{i}^2}{2k_BT_{k}}\right], ~(i=x,y)$, with which the system behaves essentially the same as an equilibrium fluid at temperature $T_{k}$. Such a behavior is independent of the density or $l_m$ (\emph{SI Appendix}, Fig. S5). In Fig.~\ref{Fig2}C, we plot the structure factor $S(q)$ of systems at $l_m=4.4\sigma$ with different $l_d$. When $l_d=96\sigma$, we find $S(q)\sim 1$, close to that of { low density} equilibrium simple fluids (open symbol).  With decreasing $l_d$, the $S(q)$ curve gradually bends down at small $q$ regime and develops a deep $q^2$ scaling, indicating a strong HU~\cite{hexner2017noise,lei2019}. Therefore, at a fixed density, the strength of inertia represented by $l_d$ determines not only how far the system is out of equilibrium, but also the length scale, above which the { hyperuniform scaling regime begins}.

\subsection*{Hydrodynamic Theory for Hyperuniform Fluids}
The random-organizing hard-sphere fluid has a well-defined kinetic temperature in the large $l_d/l_m$ limit. This motivates us to construct a fluctuating hydrodynamic theory for these hyperuniform fluids based on  Navier-Stokes (NS) equations for isothermal fluids~\cite{ramaswamy1982,gross2011model}. The modified NS equations for the local mass density $\rho$ and velocity $\mathbf u$ can be written as
\begin{eqnarray}\label{navier_stokes}
\frac{\partial \rho}{\partial t}  &=& -\nabla \cdot {(\rho \mathbf u)}, \\
\frac{\partial (\rho \mathbf u)}{\partial t} +\nabla \cdot (\rho \mathbf {u u}) &=&  -\tilde{\gamma}  \rho \mathbf u - \nabla p + \nabla \cdot (\boldsymbol{\sigma}^v + \boldsymbol{\sigma}^r), \label{navier_stokes_B}
\end{eqnarray}
where $\tilde{\gamma}=\gamma /m$ is the reduced damping coefficient and $p$ is the local pressure. We assume a simple equation of state $p=c^2_s \rho$ for the system with $c_s$ the speed of sound.  $ \boldsymbol{\sigma}^{v} $ and $ \boldsymbol{\sigma}^r $ are the classical momentum-conserved viscous stress tensor and random (noise) stress tensor for simple fluids, respectively~\cite{hansen1990theory,landau1957}. Details about the hydrodynamic theory are given in \emph{SI Appendix}.
With the additional damping term $-\tilde{\gamma}  \rho \mathbf u$, Eq.~(\ref{navier_stokes}-\ref{navier_stokes_B}) describe a non-equilibrium fluid which violates the fluctuation-dissipation theorem~\cite{ramaswamy1982}. Using the standard hydrodynamic linearization procedure, we find the density fluctuation $\delta \rho_{ \mathbf{q} }$ in $\mathbf q$ space satisfying
\begin{eqnarray} \label{dyanmic_eqt}
q^{-2} \frac{\partial^2 \delta \rho_{\mathbf{q}}}{\partial t^2}  =-(\tilde{\gamma} q^{-2} + \nu^{\parallel})  \frac{\partial \delta \rho_{\mathbf{q}}}{\partial t}  - c_s^2 \delta \rho_{\mathbf{q}} + { {\sigma}^r_{\parallel,\mathbf{q}}},
\end{eqnarray} 
with $\nu^{\parallel}$ the longitudinal kinematic viscosity.  ${\sigma}^r_{\parallel,\mathbf{q}}$ is the longitudinal component of random noise in $\mathbf q$ space which satisfies $ \langle { {\sigma}^r_{\parallel,\mathbf{q}}}(t){ {\sigma}^r_{\parallel,\mathbf{q}}}(t') \rangle = 2 \rho_0 \nu^{\parallel}k_BT_{k}V \delta(t-t')$ { with $\rho_0$ the average mass density of the system.} Eq.~(\ref{dyanmic_eqt}), in fact, describes a damped stochastic harmonic oscillator if we map $q^{-2}$ as the mass, $\delta \rho_{\mathbf{q}}$ as the displacement, $\tilde{\gamma}q^{-2} + \nu^{\parallel}$ as the effective damping coefficient, $c_s^2 \delta \rho_{\mathbf{q}}$ as the restoring force and ${ {\sigma}^r_{\parallel,\mathbf{q}}}$ as the driving noise. The average potential of the oscillator can be obtained based on the equipartition theorem~\cite{dybiec2017} if we assume the damping from a viscous thermal bath at temperature $T_{bath}$, i.e.,
\begin{eqnarray}\label{bath}
\left\langle \frac{1}{2} c_s^2 |\delta \rho_{\mathbf{q}}|^2 \right\rangle = \frac{k_BT_{bath}}{2}.
\end{eqnarray}
Here, according to the Langevin equation, $ T_{bath} =N m T_{k}\nu^{\parallel}/(\tilde{\gamma}q^{-2} + \nu^{\parallel})$. The magnitude of oscillation, or the static structure factor of the fluid, then can be written as 
\begin{eqnarray} 
S(\mathbf{q}) = \frac{1}{Nm^2} \langle |\delta {\rho}_{\mathbf{q}}|^2 \rangle = \frac{q^2 }{B l_d^{-2} + q^2},\label{Sk_final}
\end{eqnarray}
with $B=\sqrt{{8}/{A}}$ for 2D systems. Here we simply assume $ \nu^{\parallel} \simeq l_m \sqrt{k_BT_{k} /m}$ and $c_s \simeq \sqrt{k_BT_{k} /m}$ based on the kinetic theory of ideal gas, ignoring possible pre-factors in $\nu^{\parallel}$ and $c_s$. { Eq.~(\ref{Sk_final}) can be also directly obtained using our hydrodynamic theory (see \emph{SI Appendix}).} From Eq.~(\ref{bath}, \ref{Sk_final}), we find $S(q)= 1$ independent of $q$ for low density equilibrium fluids ($\tilde{\gamma}=0$), consistent with the fact that the average potential of the oscillator at a fixed temperature is independent of its mass (Fig.~\ref{Fig2}E). For non-equilibrium fluids ($\tilde{\gamma}>0$), when $ q \gg l_d^{-1}$, the $\tilde{\gamma} q^{-2}$ term in Eq.~(\ref{dyanmic_eqt}) can be neglected compared with $\nu^{\parallel}$, and the oscillator  behaves the same as when $\tilde{\gamma} = 0$. However, when $q \ll l_d^{-1}$, the $\tilde{\gamma} q^{-2}$ term dominates, leading to the decrease of oscillation magnitude according to $S(q)\simeq q^2l_d^2$ (hyperuniform scaling) independent of $l_m$. The oscillator thus has a vanishing oscillation magnitude at small $q$, i.e., $S(q\rightarrow 0)= 0$, because of the infinitely large effective damping coefficient at this point (Fig.~\ref{Fig2}F). This mechanism of fluidic HU can be also understood as that the hyperuniform fluid has a $q$-dependent temperature which vanishes at $q\rightarrow 0$. Such a special temperature property may not be captured by the non-Gaussian distribution of particles' velocity. In Fig.~\ref{Fig2}C, the results of Eq.~(\ref{Sk_final}) are shown as dashed lines, which well agree with the simulation results at small $q$.   In \emph{SI Appendix}, Fig. S6, we show the insensitivity of $S(q)$ with respect to $l_m$ for low density system far from critical point. Thus, only $l_d$ controls the length scale, above which the HU scaling occurs.  { In \emph{SI Appendix} Fig. S7, we also show how $S(q)$ changes when the system crosses the phase boundary.} We find that this $l_d$-dependency still holds when the 2D system approaches the critical point $l_d\simeq l_m$ exhibiting the critical hyperuniform scaling $S(q\rightarrow 0)\sim  q^{0.45}${~\cite{hexner2015,tjhung2015,hexner2017enhanced}}.

\begin{figure}[htbp]
	\resizebox{84mm}{!}{\includegraphics[trim=0.0in 0.0in 0.0in 0.0in]{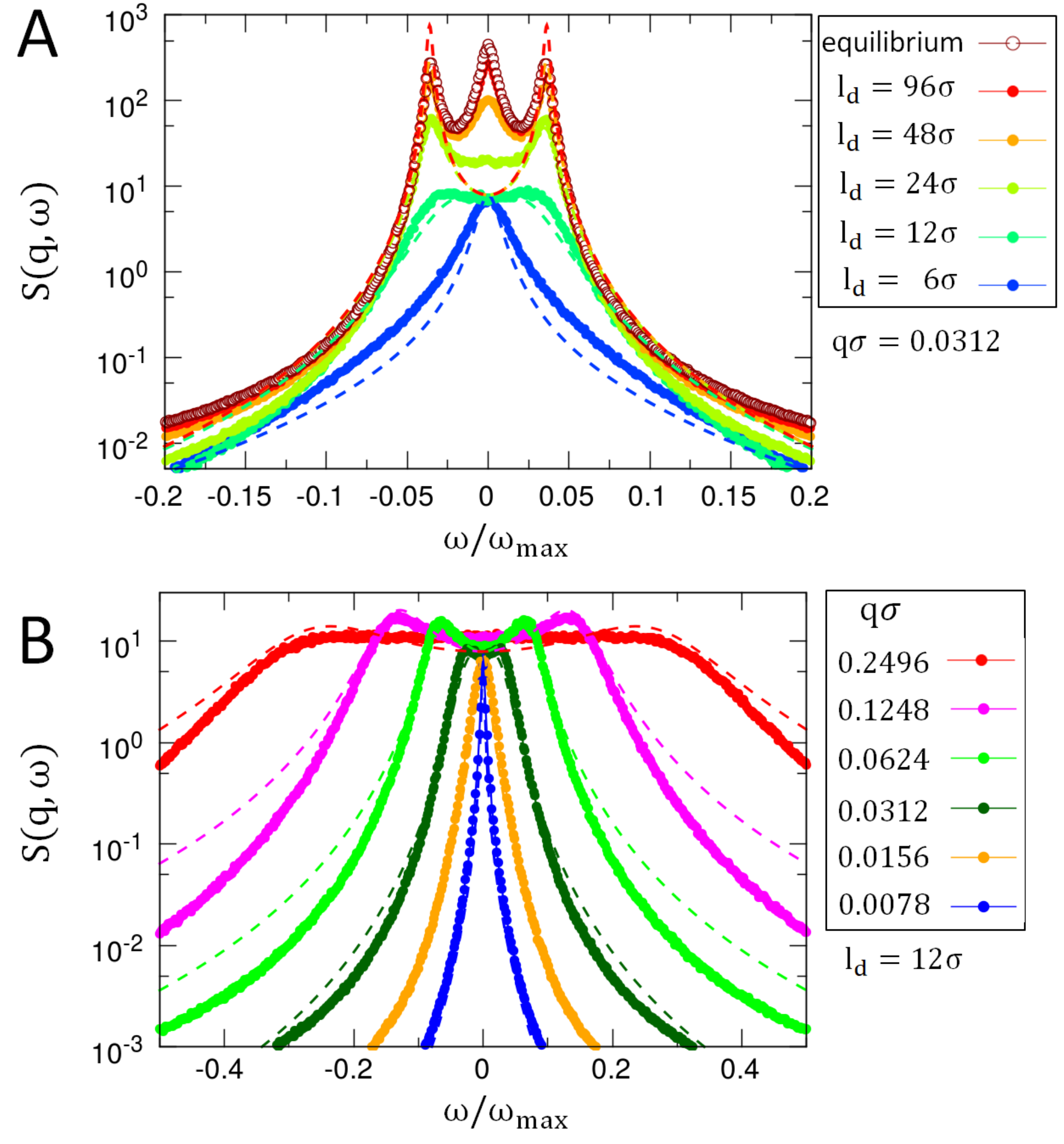} }
\caption{ Dynamic structure factor $S(q, \omega)$ for 2D systems with $q\sigma=0.0312$ at different $l_d$ ({\emph{A}}),  and for systems with $l_d=12\sigma$ at different $q$ (\emph{B}). For all cases, $\tilde{\rho}=0.1$ ($l_m=4.4\sigma$). The dashed lines show the theoretical predictions of Eq.~(\ref{dynamic_structure_factor}) multipled by an adjustable coefficient 1.9. Here, $\omega_{max}=2\pi c_s/l_m$.}
\label{Fig3}
\end{figure} 

Dynamically, the oscillator represented by Eq.~(\ref{dyanmic_eqt}) is in the resonance state in regime:
\begin{eqnarray}
 \frac{c_s-\sqrt{c_s^2-\tilde{\gamma} \nu^{\parallel} }}{\nu^{\parallel}} < q  <  \frac{c_s+\sqrt{c_s^2-\tilde{\gamma} \nu^{\parallel} }}{\nu^{\parallel}}. \label{sound_region}
 \end{eqnarray}
This means that the fluids can support acoustic (sound) modes~\cite{ramaswamy1982}. Due to the damping from $\gamma$, these sound modes are fast modes, which decay as $e^{-\kappa t}$ with damping rate $\kappa \simeq \frac{1}{2} (\tilde{\gamma} + \nu^{\parallel}q^2)$. For low density systems with $l_d\gg l_m$, this acoustic regime can be approximated as $\frac{B l_m}{2}  l_d^{-2} < q  <2 l_m^{-1}$. With decreasing $q$ or increasing $\gamma$, the oscillator crosses from the resonance regime to the overdamped regime, in which the system only has the diffusive mode $\omega^*_{1} \simeq -iDq^2$ with the diffusion constant $D=c_s^2/\tilde{\gamma}$.  In fact, in this overdamping limit, Eq.~(\ref{dyanmic_eqt}) can be written as a simple diffusion equation,
\begin{eqnarray} \label{dyanmic_eqt2}
\frac{\partial \delta \rho_{\mathbf{q}}}{\partial t} =  - D q^{2} \delta \rho_{\mathbf{q}} + q^{2} { {\sigma}^r_{\parallel,\mathbf{q}}}.
\end{eqnarray} 
This equation was employed to explain the HU in a time-discrete random-organizing model with  {the `center of mass conservation', a discrete form of reciprocal interaction in overdamped system, which breaks down in continuous space-time when particles have finite inertia~\cite{hexner2017noise}.}   In Fig.~\ref{Fig2}G, we plot the schematic diagram of density modes in the representation of $1/l_d$ and $q$. One can see that when $l_d<l_m$, the system stays in an absorbing state without any dynamic mode (blue region), while for $l_d>l_m$, the system has the diffusive mode (red region) and acoustic modes (green region) at relative small $q${, and non-acoustic `fast-decay mode' at large $q$.} The upper boundary of the acoustic regime is determined by Eq.~(\ref{sound_region}), which becomes less accurate (dashed line) when the system approaches the critical point $l_d\simeq l_m$. The shadow area with bottom boundary $q \sim l_d^{-1}$ represents the $q$ space regime exhibiting the $q^2$ hyperuniform scaling. This area is mainly located at the diffusive regime, but also overlaps with the acoustic regime in ${l_m} l_d^{-2} < q \ll l_d^{-1}$, at which  $S(q)\simeq q^2l_d^2\rightarrow0$ for $l_d\rightarrow\infty$. Therefore, the suppressed long-wavelength density fluctuation  in hyperuniform fluids can exhibit as either  diffusive  or acoustic modes, { though at \emph{infinite} wavelength ($q=0$), only diffusive mode exists. }

{  Generally, different dynamic modes can be characterized by different shapes of dynamic structure $S(q,\omega)$ with $\omega$ the angular frequency. For acoustic modes, there would be two symmetric Brillouin peaks $\omega_{\pm}=\pm c_s q$ in $S(q,\omega)$, while for diffusive modes, only a single peak exists round $\omega=0$. Our hydrodynamic theory predicts
\begin{eqnarray}\label{dynamic_structure_factor} 
S(\mathbf{q},\omega) &=&   \frac{ 2q^4 \nu^{\parallel} m^{-1} k_BT_{k}}{(\omega^2 - c_s^2 q^2)^2 + \omega^2(\tilde{\gamma}  + \nu^{\parallel}q^2)^2}.
\end{eqnarray}
In Fig.~\ref{Fig3}, we show the measured $S( q,\omega)$ in simulation as a function of $\omega$ (symbols) for systems with fixed $l_d$ but different $q$ (Fig.~\ref{Fig3}A), and systems with fixed $q$ but different $l_d$ (Fig.~\ref{Fig3}B) at the same density $\tilde{\rho}=0.1$. One can clearly distinguish the acoustic modes in large $l_d$ (or $q$) cases and diffusive modes in small $l_d$ (or $q$) cases, respectively, in Fig.~\ref{Fig3}A (or Fig.~\ref{Fig3}B). The theoretical predictions of Eq.~(\ref{dynamic_structure_factor}) are plotted as dashed lines by using the accurate expression of sound speed $c_s^a$ for the equilibrium fluid, which takes into account the effects  of the adiabatic compression and finite density of system, i.e., $c_s^a=\sqrt{\Gamma/\rho_0 \chi_{T}}$, with $\Gamma=1+2/d$ the adiabatic index for monoatomic fluids and $\chi_{T}=S_{\rm eq}(0)m(\rho_0 k_BT_k)^{-1}$ the isothermal compressibility~\cite{hansen1990theory}. Here, $S_{\rm eq}(0)=0.72$ is obtained from the equilibrium system in Fig.~\ref{Fig2}C. We find that the theoretical predictions of Brillouin peaks match well not only with data from the equilibrium fluid but also with non-equilibrium hyperuniform fluids. This indicates that the sound speed and the compressibility in hyperuniform fluids with kinetic temperature $T_k$ should be close to that of the the equilibrium system with the same $T_k$ despite their huge difference in $S(0)$. For equilibrium fluids (energy conserved) or near-equilibrium fluids, due to the heat diffusion, the measured $S(q,\omega)$ shows an additional Rayleigh peak at $\omega=0$~\cite{chaikin2000principles,hansen1990theory}. Although our isothermal hydrodynamic theory without energy conservation can not capture this feature (see {\emph{SI Appendix}} for the discussion), it predicts the bimodal acoustic modes and diffusive modes faithfully for hyperuniform fluids with small $l_d/l_m$ ratio where the heat diffusion is negligible. Especially, in the overdamping limit, both the simulation and theoretical results in Fig.~3B indicate $S(q,\omega)\rightarrow const.~(\omega \rightarrow 0)$, i.e., a fixed point for hyperuniform fluids, while for non-hyperuniform equilibrium Brownian particle system, $S(q,\omega)\propto \frac{q^2}{\omega^2 +D^2q^4} \rightarrow  (Dq)^{-2}~(\omega \rightarrow 0)$, i.e., a standard Lorentz distribution~\cite{chaikin2000principles}. The above dynamic structure features  of random-organizing hyperuniform fluids facilitate their experimental detection using classical light scattering methods~\cite{chaikin2000principles,hansen1990theory}. }

\begin{figure*}[!htb]
\centering
\begin{tabular}{c}
	\resizebox{175mm}{!}{\includegraphics[trim=0.0in 0.0in 0.0in 0.0in]{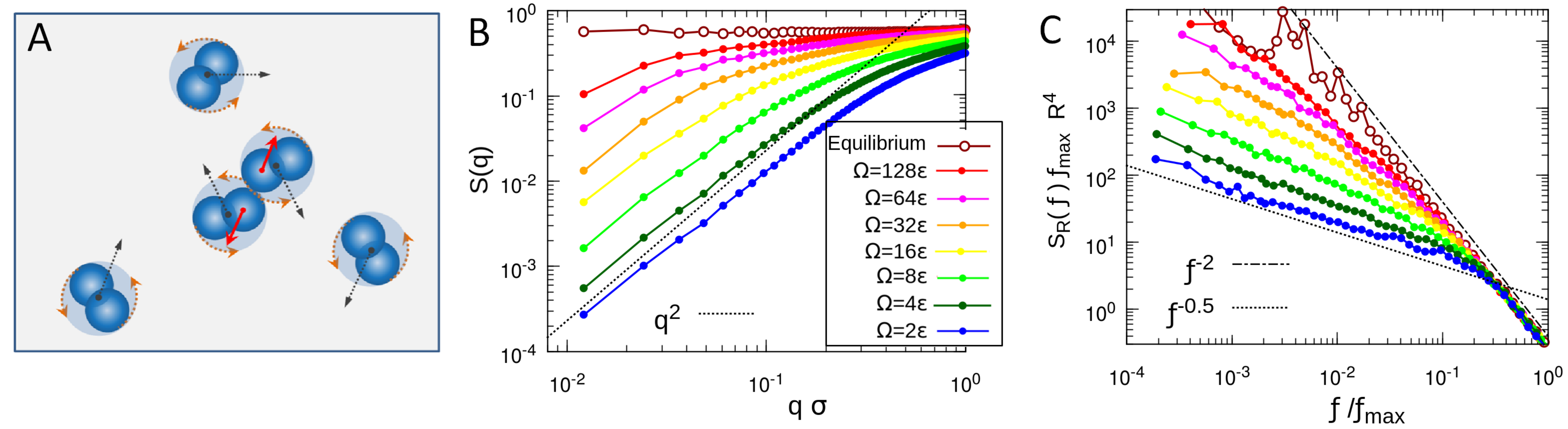} }
\end{tabular}
\caption{({\emph A}) Schematic of 2D active spinners. (\emph{ B, C}) Structure factor $S({q})$ and power spectrum $S_R(f)$ for systems with different $\Omega$ at $\tilde{\rho}=0.15$, where $f_{max}=\sigma\sqrt{m/k_BT_{k}}$. The open symbols show the data for the equilibrium dimer fluid. For all the simulations, $N=40,000$. }
\label{Fig4}
\end{figure*} 

\subsection*{Power Spectrum of Local Density Fluctuations}
The aforementioned dynamic regimes in hyperuniform fluids can also be distinguished by the noise color reflected by the power spectrum of local density fluctuations ~\cite{jensen1998self,fogedby1991temporal}, i.e.,
\begin{eqnarray}
S_R(f)\equiv \frac{1}{m^2\tau_{max}}\left|\int^{\tau_{max}}_0 \rho_R(t)e^{2i\pi ft}dt\right|^2,
\end{eqnarray}
where $f=\omega/2\pi$ is the frequency and {$\tau_{max}$ is the maximum observation time.} $\rho_R(t) \equiv m N_R(t)/\pi R^2$ is the local mass density with $N_R(t)$ the number of particles in a circular sub-domain of radius $R$. The calculation details and the complete expression of $S_R(f)$ based on our hydrodynamic theory can be found in \emph{SI Appendix}. Here, we only summarize the main results. We first introduce three typical frequencies $\tilde{\gamma}$, $f_{1}$ and $f_{max}$, where $\tilde{\gamma}^{-1}$ is the typical dissipation time, $f_1^{-1}=R/c_s$ is the time with which the acoustic wave propagates a distance $R$ and $f_{max}=c_s/l_m$ is the typical cutoff frequency for the acoustic wave. Here, $\tilde{\gamma}/f_{max} \propto ({l_m}/ {l_d})^2$. For equilibrium simple fluids ($\tilde{\gamma}/f_{max} \rightarrow 0$), or more generally, $ \mathrm{max}\left[f_1, (\tilde{\gamma} f_{max})^{1/2}\right]  \ll  f \ll f_{max} $, the power spectrum exhibits a `red' noise
\begin{eqnarray} 
S_R(f) ~\sim ~f^{-2}.
\end{eqnarray}
For overdamped hyperuniform fluids ($\tilde{\gamma}/f_{max} \sim 1$), or more generally, $\mathrm{min}(f_1,f_1^2/\tilde{\gamma} ) \ll f \ll \tilde{\gamma}$, we have
\begin{eqnarray} 
S_R(f)\sim ~f^{-1/2}.
\end{eqnarray}
This special noise is different from the $f^{-3/2}$ noise found in non-hyperuniform equilibrium Brownian particle systems~\cite{jensen1998self,fogedby1991temporal}. Moreover, there is also a crossover regime, i.e., $\tilde{\gamma} \ll f \ll (\tilde{\gamma} f_{max})^{1/2}$, in which the power spectrum behaves as `white' noise
\begin{eqnarray} 
S_R(f)\sim f^{0}.
\end{eqnarray}
In Fig.~\ref{Fig2}D, we plot $S_R(f)$ measured in the simulation for the same systems as shown in Fig.~\ref{Fig2}C but distinguish them by using $\tilde{\gamma}/f_{max}$. The theoretical predictions from our hydrodynamic theory multiplied by an adjustable coefficient $0.62$ are shown as dashed lines.  We find that when $\tilde{\gamma}/f_{max}$ is small, the measured $S_R(f)$ shows the `red' noise scaling in a broad range of frequency accompanied with some acoustic peaks around $f_1$.  With increasing $\tilde{\gamma}/f_{max}$, the acoustic peaks fade away and the special $f^{-1/2}$ noise scaling gradually dominates for $f\ll \tilde{\gamma}$. The white noise plateau can only be seen for the cases of $\tilde{\gamma} /f_{max}=1.5\sim 6.0\times 10^{-2}$  due to the finite size of our system.  Comparing Fig~\ref{Fig2}C with Fig~\ref{Fig2}D, we find that the development of hyperuniform scaling $q^2$ in $S(q)$ with decreasing $l_d$ is associated with the suppression of temporal fluctuation at small frequency regime with increasing $\tilde{\gamma}/ f_{max}$. Actually, the `red' noise regime corresponds to the non-overlapped acoustic region in Fig.~\ref{Fig2}G, while the white noise regime with upper-bound frequency $(\tilde{\gamma} f_{max})^{1/2} \sim c_s/l_d$ and the $f^{-1/2}$ noise regime with the upper-bound frequency $\tilde{\gamma}\sim c_s l_m/l_d^2$ correspond to the acoustic and diffusive hyperuniform regions in Fig.~\ref{Fig2}G, respectively. These special noise features provide us with an additional tool to identify and study the hyperuniform fluids experimentally in the frequency domain. We emphasize that all these structural and dynamic properties of our hyperuniform fluids are independent of dimensionality and we show the simulation data for 3D system in \emph{SI Appendix}, Fig. S8, S9 and Movie S4.  

\subsection*{Hyperuniform Active Spinner Fluids}
The hydrodynamic theory above suggests two important conditions for forming non-equilibrium hyperuniform fluids: i) the damping dissipation characterized by $\tilde{\gamma}$, which has a dominant effect at small $q$, and ii) the active reciprocal interaction (driving), which conserves the classical formula of viscosity and the random noise. Based on this, we further consider a realistic system of $N$ active spinners on a frictional substrate~(Fig.~\ref{Fig4}A) as an example to realize the hyperuniform fluid. Active spinners systems exhibit many interesting phenomena and have been extensively studied  both theoretically~\cite{nguyen2014,van2016spin,banerjee2017,souslov2019} and experimentally~\cite{farhadi2018,scholz2018inertial,kokot2017,scholz2018r,
sabrina2018shape,shields2018}. We employ the active spinner model from Ref.~\cite{van2016spin} (also see \emph{SI Appendix} for details), in which each spinner is a dimer consisting of two spherical hard monomers. Driven by a constant  torque $\Omega$, spinners perform self-rotating motion with the same chirality.  The dynamics of both translational and rotational freedom degrees of spinners are underdamped.  Thus, when two fast-rotating spinners collide, some part of their rotational kinetic energy is transferred to their translational freedom degree, inducing the active collision similar to the previous hard sphere model. { With the fixed substrate friction coefficient}, we find $\Omega$ in the spinners system plays the similar role of $\Delta E$ in the previous model to control the phase behaviour of the system. A larger $\Omega$ results in a faster rotation of spinners, which leads to more energy transfer in active collisions. In Fig.~\ref{Fig4}B and C, we calculate $S(q)$ and $S_R(f)$ for the active spinners system at $\tilde{\rho}=0.15$ with different $\Omega$ (solid symbols) compared with the equilibrium dimer fluid (open symbols).  When $\Omega$ is small, we find the same hyperuniform scaling  $S(q\rightarrow 0)\sim q^2$ and $S_R(f \rightarrow 0)\sim f^{-1/2}$ in the spatial and temporal domains as previously found. The response of $S(q)$ and $S_R(f)$ to the increase of $\Omega$ is also similar to the response to $\Delta E$ in the previous model. Nevertheless, we emphasize that in the active spinner system the interacting force is not center to center and the energy $\Delta E$ is not a constant. These differences demonstrate the robustness of fluidic HU, but also request a modified hydrodynamic theory for active spinner fluids to describe the `odd viscosity' and topological waves in the fluids~\cite{banerjee2017,souslov2019}. Finally, we note that the suppressed density fluctuation was detected in a recent experimental air-driven spinner system~\cite{farhadi2018} and we expect further experimental realization of hyperuniform spinner fluids using different techniques~\cite{scholz2018inertial,kokot2017,scholz2018r,
sabrina2018shape,shields2018}.

\section*{Discussion and Conclusion}
In conclusion, a general hydrodynamic mechanism of fluidic HU is discovered by investigating a random-organizing hard-sphere fluid model and formulating a fluctuating hydrodynamic theory based on the generalized Navier-Stokes equations. We find that the state of the system is determined by two characteristic lengths: the dissipation length $l_d$, reflecting the strength of the particle inertia, and the mean free path $l_{m}$ associated with the particle density. A smooth transition from an equilibrium hard sphere fluid to a non-equilibrium hyperuniform fluid occurs with decreasing $l_d$ from infinity to $l_m$, below which the system falls into an absorbing state. We determine this absorbing phase transition belonging to the universality of conserved directed percolation (CDP) for systems with low particle inertia ($l_d$). Increasing the inertia induces the crossover of the universality class from CDP to its long-range mean-field scenario. These hyperuniform fluids  have a kinetic temperature controlled by ratio $l_d/l_m$, which reflects how far the system is out of equilibrium, while $l_d$ alone sets the length scale, above which the hyperuniform scaling occurs. Importantly, we demonstrate that the mechanism of fluidic HU can be understood as the damping of a $q$-space stochastic harmonic oscillator, and the suppressed long-wavelength density fluctuation in the fluids exhibits as either acoustic (resonance) or diffusive (overdamped) modes. These dynamic modes are identified in the dynamic structure factor and the power spectrum of local density fluctuations. Our theory suggests two important ingredients for the random-organizing fluidic HU, i.e., (under/over)damped dynamics (dissipation) and  reciprocal active interaction (driving).  Accordingly, we show the fluidic HU with controllable hyperuniform length scales in an experimentally realizable active spinner system.  Generally, the damping can be induced by the substrate friction, viscous solvent/gas or `optical molasses' (laser cooling)~\cite{chu1985three}, while the reciprocal active interaction can be realized through energy transfer between different degrees of freedom~\cite{scholz2018r,castillo2019h}, active deformation~\cite{li2019,zehnder2015cell}, localized motion~\cite{lei2019,han2017effective}, time-oscillatory potentials~\cite{t2014dissipative} or dielectric heating~\cite{grant1998dielectric}. Thus, we anticipate that other driven-dissipative systems studied previously can also produce the fluidic HU, e.g., the vibrated granular monolayer which allows the kinetic energy transferring from vertical to horizontal freedom degree~\cite{castillo2019h}, actively deformable micro-robots~\cite{li2019}/cells~\cite{zehnder2015cell} on a substrate or in solvent, or molecules gas in optical molasses~\cite{shuman2010laser} with vibrational/rotational excitation~\cite{grant1998dielectric,v2000forced}. { These  dynamic hyperuniform fluids, if realized, would be capable of self-healing from damage and self-adapting to the change of external driving. Thus, compared with solid hyperuniform materials~\cite{donevprl2005,florescu2009designer,man2013isotropic,man2013photonic}, they behave essentially like a living functional material~\cite{bachelard2017emergence},  which may have  potential applications in optics, acoustics and micro-fluidic engineering.}

\matmethods{} \showmatmethods{}
{In our simulations, we use square (2D) or cubic (3D) simulation boxes to calculate the $S(q)$ and $S_R(f)$, while in calculate $S(q,\omega)$ elongated boxes with aspect ratio $1:8$ (2D) and $1:1:8$ (3D) are used. The mean free path $l_m(\tilde{\rho})$ in the hard sphere model is calculated using randomly-generated static configurations to ensure that $l_m$ is only a function of $\tilde{\rho}$ and independent of the dynamic state of the system.  We have further checked that HU has no significant influence on $l_m(\tilde{\rho})$. In the calculation of $l_m$, particles' velocities are all set to zero. We then give a randomly-selected particle $i$ a random direction and obtain the maximum distance the particle can travel in a straight line (without damping) before colliding with another particle $j$. Then, the particle $i$ is put back to its original position and particle $j$ is selected as the next moving particle. The structure factor {is estimated by the scattering intensity} 
$S({\mathbf q})=\frac{1}{N} \left< \left|\sum_{j=0}^{N} e^{i{\mathbf q}\cdot{\mathbf r}_j} \right|^2\right>$ with ${\mathbf q}=[q_{x}, q_{y}, q_{z}]=[i\frac{2\pi}{L}, j\frac{2\pi}{L}, k\frac{2\pi}{L} ]$ ( $i,j,k =1,2,3, \cdots$) and $L$ the box size. The vector $\mathbf q$ is then projected to the scalar $q$. For 2D systems, $k=0$. The  dynamic structure factor {is estimated as} $S({\mathbf q},\omega)=\frac{1}{N\tau_{max}} \left|\sum_{j=0}^{N} \int_0^{\tau_{max}} e^{i[{\mathbf q}\cdot{\mathbf r}_j(t)-\omega t] } dt \right|^2$ with angular frequency  $\omega =2\pi j/\tau_{max}$ ($j =1,2,3, \cdots$). The direction of $\mathbf q$ is chosen to be along with the elongated box. In the calculation of $S_R(f)$, the frequencies are chosen similarly and the sub-domain radius is set to $R=L/4$. To increase the efficiency, we sample 100 random-located sub-domains at the same time.}

Details about the event-driven simulations, finite-size scaling analysis of absorbing phase transitions, hydrodynamic theory, theoretical derivations of $S_R(f)$ in 2D and 3D, the molecular dynamic simulation of active spinners are included in the \emph{SI Appendix} Supplementary text. {A table list of symbols used in this work can be found in \emph{SI Appendix} Table S2. The original data for Fig.~1-4 is provided in \emph{SI Appendix} Database S1-S9}.

\acknow{
The authors are grateful to Profs. Dov Levine and Hao Hu for helpful discussions. This work is supported by Nanyang Technological University Start-Up Grant (NTU-SUG: M4081781.120), the Academic Research Fund from Singapore Ministry of Education (M4011873.120), and the Advanced Manufacturing and Engineering Young Individual Research Grant (A1784C0018) by the Science and Engineering Research Council of Agency for Science, Technology and Research Singapore.}

\showacknow{}

%

\bibliography{reference}

\end{document}